# Giant Magnetoimpedance of Glass-Covered Amorphous Microwires of Co-Mn-Si-B and Co-Si-B

H. B. Nie, X. X. Zhang, A. B. Pakhomov, Z. Xie and X. Yan

*Department of Physics, Hong Kong University of Science and Technology, Clear Water Bay, Kowloon, Hong Kong*

A. Zhukov and M. Vázquez

*Instituto de Magnetismo Aplicado, P.O. Box 155, 28230, Las Rozas, Madrid, Spain*

Abstract

A study of magnetic hysteresis and Giant magnetoimpedance (GMI) in amorphous glass covered Co-Si-B and Co-Mn-Si-B wires is presented. The wires, about 10 microns in diameter, were obtained by glass-coated melt spinning technique. Samples with positive magnetostriction (MS) have a rectangular bistable hysteresis loop. A smooth hysteresis loop is observed for wires with nearly zero MS. When MS is negative, almost no hysteresis is observed. The GMI was measured in the frequency range between 20 Hz and 30 MHz. The shapes of the impedance versus field curves are qualitatively similar to each other for both positive and zero MS samples. Impedance is maximum at zero field, and decreases sharply in the range 10-20 Oe. For the negative MS wires, when the driving current is small, the impedance is maximum at a finite external field. The position of the maximum approaches zero with increasing current. The contributions of the moment rotation and domain wall motion in the three cases are discussed.

The giant magnetoimpedance (GMI)[1] in conducting soft magnetic materials such as amorphous or nanocrystalline wires, ribbons and films is considered as a base for high frequency magnetic sensor applications. It is established that the GMI can be associated with the effect of the applied dc magnetic field on the magnetic permeability in the direction transverse with respect to the ac current through the conductor, or, for cylindrical samples, in the circumferential direction. For a cylindrical magnetic conductor, the impedance can be written in the form:[1] $Z = R_{dc} ka J_0(ka)/2J_1(ka)$, where $k = (1+i)/\delta_m$, $\delta_m = c/(2\pi\omega\sigma\mu_\phi)^{1/2}$, $R_{dc}$ is the dc resistance, $J_0$ and $J_1$ are the Bessel functions, $a$ is the radius of the wire, $\delta_m$ is the penetration depth, $\omega = 2\pi f$ is the frequency, $\sigma$ the conductivity and $\mu_\phi$ the effective complex magnetic permeability in the circumferential direction. The dominating contribution to the effective permeability can be due either to the rotation of magnetization or to domain wall motion, depending on the magnetic anisotropy and domain wall stiffness. In the amorphous soft magnetic materials, the sign and magnitude of magnetostriction (MS), combined with the type of stresses in the samples, play a dominant role in the formation of the magnetic domain structure and hence in the sensitivity of the GMI effect. Glass-covered soft magnetic amorphous microwires[2] constitute an important class of materials of potential importance for applications. GMI has been studied, for example, in CoFeSiB and FeCuNbSiB glass covered wires.[3,4]. In this article we report measurements of the GMI in amorphous $Co_{75}Si_{10}B_{15}$, $Co_{68.2}Mn_{6.8}Si_{10}B_{15}$, and $Co_{68}Mn_{7}Si_{10}B_{15}$ glass covered wires. The values of magnetostriction in these three materials are respectively negative, nearly zero and positive, thus providing a possibility of a comparative study of GMI in samples with three different types of magnetic anisotropy.

The wires, with core diameters between 8 μm and 12.5 μm and the glass cover thickness between 4.3 μm and 7.5 μm, were obtained by glass-coated melt spinning technique.[2] The dc hysteresis loops were measured by a superconducting quantum interference device (SQUID) magnetometer in the direction parallel to the sample axes. The GMI was measured on a HP 4284A *RLC* meter in the frequency range between 20 Hz and 1 MHz, and HP 4285A *RLC* meter in the range 75 kHz- 30 MHz. A Helmholtz coil was used to generate magnetic field parallel to the wire axis in the range $-115.5$ Oe $\leq H \leq$ 115.5 Oe.

Figure 1 shows the measured magnetization curves for the three samples. The wire of the base alloy $Co_{75}Si_{10}B_{15}$ has a negative MS.[5] A loop with almost no hysteresis is observed for this sample [Fig. 1(a), curve 1]. This behavior has been attributed[5] to circular domain structure due to magnetoelastic anisotropy. It is related to the distribution of stressed generated in the wire preparation process and is due to the difference in the coefficients of thermal expansion of metal and glass. Curve 2 in Fig. 1(a) shows the hysteresis loop for this sample after annealing by ac current with a root mean square (rms) value of 50 mA for 0.5 hours. After annealing the anisotropy field $H_K$ is reduced to about 50 Oe compared to 80 Oe in the as-prepared sample. The substitution of cobalt atoms by manganese leads to a decrease of the absolute value of negative MS. As has been reported before,[6] in $Co_{68.5}Mn_{6.5}Si_{10}B_{15}$ the hysteresis loop is very similar to that shown in Fig. 1(a), except that the value of anisotropy field $H_K \approx$ 4 Oe. Close to this composition, even a very small increase of the Mn content drastically changes the magnetic behavior of the samples.

A smooth hysteresis loop shown in Fig. 1(b) is observed for wires of $Co_{68.2}Mn_{6.8}Si_{10}B_{15}$. According to the analysis of Ref. 2, this behavior is a signature a MS very close to zero. One can assume that a magnetic structure resulting in the loop of Fig. 1(b) may be a combination of domains directed along the sample axis in the metal wire core and radial magnetization in the outer shell close to the metal-glass interface. The picture is very different for the sample of $Co_{68}Mn_{7}Si_{10}B_{15}$ (Fig. 1(c)). This sample has a sharp rectangular hysteresis loop associated with positive magnetostriction[2,5] and axial alignment of magnetization. An almost square and very narrow loop, with the coercive force $H_C$ smaller than 1 Oe, shows that this sample has a quasi single domain magnetic structure where the domain wall is easily moved by the external axial field.

Figure 2 shows the dependencies of the modulus of impedance $Z = |R+iX|$, at several frequencies, on the axial external magnetic field $H$ for the same three samples. We will start our consideration with the samples with zero and positive MS. One can see that the shapes of the dependencies for samples with radial and axial magnetizations (Fig. 2(b) and 2(c)) are almost similar to each other. The impedance has a sharp maximum at zero field, and saturates to a constant value depending on frequency in the field of about 25 Oe. At a frequency of 30 MHz, the saturated MI ratios $\Delta Z/Z_s$ for these two samples are about 153% and 138 %, where $\Delta Z/Z_s$ is defined as $[Z(H = 0) - Z(H = 115\text{ Oe})]/Z(H = 115\text{ Oe})$. The GMI has a maximum sensitivity $\left| [d(\frac{\Delta Z}{Z}) / dH] \right|_{\max}$ of 65 %/Oe for sample $Co_{68.2}Mn_{6.8}Si_{10}B_{15}$, and 66 %/Oe for sample $Co_{68}Mn_7Si_{10}B_{15}$. The field and frequency dependences of the impedance for these samples are in line with the conventional interpretation of GMI in terms of classical electrodynamics,[1] assuming that the soft magnetic properties are due to magnetic moment rotation.

The dependencies of the impedance of the as-cast sample with composition $Co_{75}Si_{10}B_{15}$ on the axial magnetic field $H$, at a driving current of 2.5 mA and frequencies 1, 15 and 30MHz, are shown in Fig. 2 (a). Figures 3(a) and 3(b) show the behavior for the same sample before and after annealing respectively, at a fixed frequency f = 1 MHz. At a small current, the impedance as a function of the axial field has two peaks at the peak values of the field $\pm H_P$. The value of $H_P$ increases slowly (approximately logarithmically) with frequency, from 30 Oe at f = 10 kHz to about 60 Oe at 30 MHz. This GMI behavior is qualitatively similar to that reported previously for a wire with a much smaller negative MS.[6] One can also notice that when the current is small, at low frequency $f \leq 1$ MHz there is a local maximum at zero field, with the width of several Oe [Fig. 3(a)]. At higher frequency this maximum is replaced by a minimum as seen in Fig. 2(a). Figure 3 shows the plots of normalized impedance of this sample versus external longitudinal magnetic field at different values of the driving current, at a frequency of 1 MHz, both for the as-cast sample [Fig. 3(a)] and the sample after annealing by the ac current of 50 mA for 0.5 hours [Fig. 3(b)]. With increasing current, the peak field $H_P$ decreases and the height of the peaks at $H_P$ increases. The two maxima merge into a single peak when the current exceeds the critical value of $I_C \approx 20$ mA, which corresponds to a circumferential field $H_C \approx 6.7$ Oe. After annealing, both the values of the peak field and the critical current are smaller than those in the as-cast sample.

We assume that there are three physically different contributions to the GMI behavior in sample $Co_{75}Si_{10}B_{15}$ related to circumferential anisotropy. The low-current low-field behavior of the magnetoimpedance can be understood as being due to the field and frequency dependence of cicumferential magnetic permeability and hence of the skin depth. The reversible motion of domain walls about their equilibrium positions is dominant when the frequency of the driving current is below the domain wall relaxation frequency $f_{dw} \sim 1$ MHz.[7] This part of circumferential permeability is strongly suppressed by the axial magnetic field,[7] and hence leads to a maximum of $Z$ at zero field. At high frequency, the rotational component becomes more important which increases with field in small fields providing a minimum of $Z$ at $H = 0$.

The maxima centered at larger fields 30 Oe<H<60 Oe, which exist at all frequencies, are difficult to be explained in the above model. It is likely that these features are not directly related to the skin depth. Instead, one can use the explanation proposed in Ref. 8 for the GMI in nonmagnetic wires electroplated with a magnetic layer with circumferential anisotropy. The voltage along the axis of the quasisingle domain cylindrical film is associated, via the Faraday's law, with a flux jump during a domain wall motion under action of the current-induced circumferential magnetic field. The apparent features of the effect are similar to those observed in our sample. However, it was shown[8] using the formalism of the Stoner-Wolhfarth model,[9] that the cicrcumferential field needed to merge the two peaks of GMI into one is close to the anisotropy field $H_K$. In our sample, the critical circumferential field $H_C \approx 6.7$ Oe is considerably smaller than $H_K \approx 80$ Oe (for the as-prepared sample). This may be due to some anisotropy dispersion associated with the stress inhomogeneity.

One of the authors (X. X. Zhang) would like to thank the RGC of Hong Kong (HKUST61111/98P and DAG97/98P) for support.

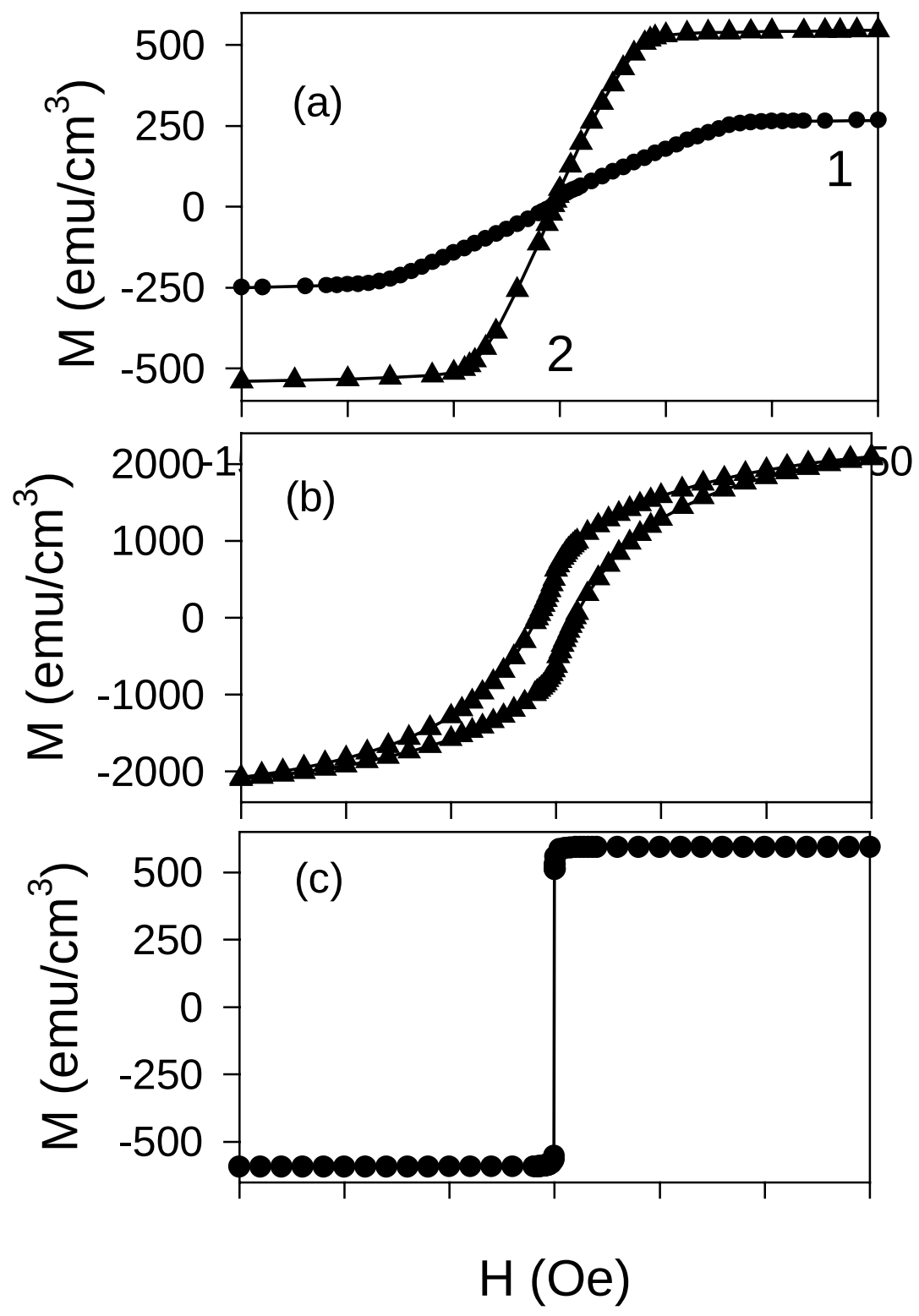

Fig. 1. Axial hysteresis loops for samples with compositions $Co_{75}Si_{10}B_{15}$ (a), $Co_{68.2}Mn_{6.8}Si_{10}B_{15}$ (b) and $Co_{68}Mn_7Si_{10}B_{15}$ (c). In panel (a) curve 1 is the hysteresis loop for as-cast amorphous wire, and curve 2 is for the same sample after annealing by ac current of 50 mA for 0.5 hours.

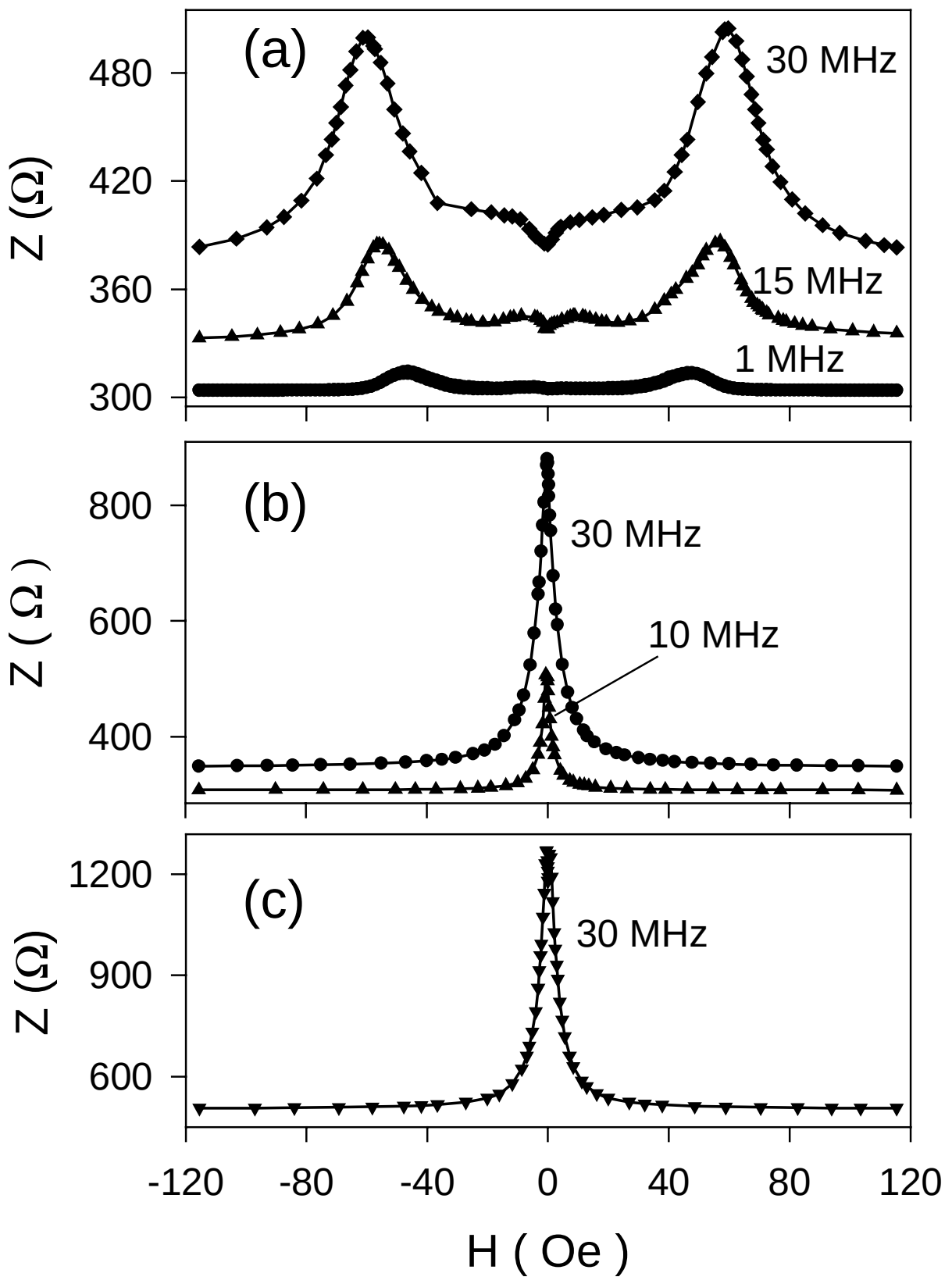

Fig. 2. GMI for samples of $Co_{75}Si_{10}B_{15}$ (a), $Co_{68.2}Mn_{6.8}Si_{10}B_{15}$ (b) and $Co_{68}Mn_7Si_{10}B_{15}$ (c), at a driving current of 2.5 mA. The values of the current frequency are shown.

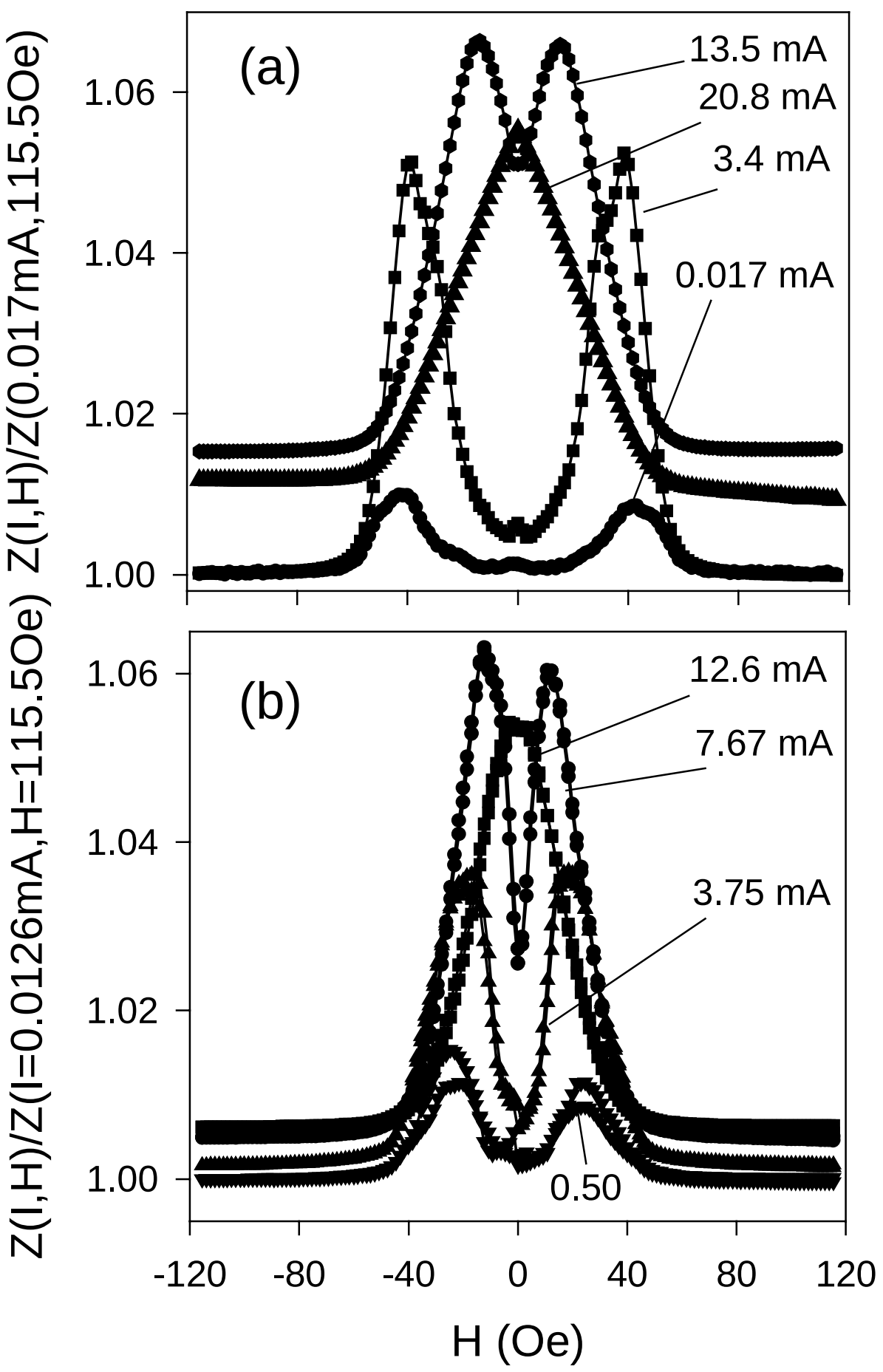

Fig. 3. Normalized GMI Z/$Z_s$, where the saturated value at small current is $Z_s = Z$ ($I$ = 0.02 mA, $H$ = 115.5 Oe), for sample $Co_{75}Si_{10}B_{15}$ at 1 MHz, at different driving current amplitudes; (a) as-cast, (b) after annealing by ac current.